\documentclass[twocolumn,showpacs,aps,prl]{revtex4-1}

\usepackage{amssymb}
\usepackage{graphicx}
\usepackage{dcolumn}
\usepackage{bm}
\usepackage{epsfig}
\usepackage{epstopdf}
\usepackage{amsfonts}
\usepackage{keyval,graphicx}
\usepackage{textcomp,wasysym}
\usepackage{subfigure}
\usepackage{hyphenat}
\usepackage{xcolor}
\usepackage{multirow}

\long\def\inst#1{\par\nobreak\kern 4pt\nobreak
    {\itshape #1}\par\vskip 10pt plus 3pt minus 3pt}
\RequirePackage{xspace}
\usepackage{relsize}

\begin{document}
\newcommand{\Dp}{D^{+}}
\newcommand{\Dm}{D^{-}}
\newcommand{\Dz}{D^{0}}
\newcommand{\Dzb}{\bar{D^{0}}}
\newcommand{\pip}{\pi^{+}}
\newcommand{\pim}{\pi^{-}}
\newcommand{\piz}{\pi^{0}}
\newcommand{\Dstp}{D^{*+}}
\newcommand{\Dstm}{D^{*-}}
\newcommand{\Dstz}{\bar{D}^{*0}}

\newcommand{\ar}{\rightarrow}
\newcommand{\GeV}{GeV/$c^2$}
\newcommand{\MeV}{MeV/$c^2$}
\newcommand{\br}[1]{\mathcal{B}(#1)}
\newcommand{\Br}{\mathcal{B}}
\newcommand{\zcn}{Z_{c}(3885)^{0}}
\newcommand{\zcpm}{Z_{c}(3885)^{\pm}}
\newcommand{\zcc}{Z_{c}(3885)^{+}}
\newcommand{\cinst}[2]{$^{\mathrm{#1}}$~#2\par}
\newcommand{\crefi}[1]{$^{\mathrm{#1}}$}
\newcommand{\crefii}[2]{$^{\mathrm{#1,#2}}$}
\newcommand{\crefiii}[3]{$^{\mathrm{#1,#2,#3}}$}
\newcommand{\HRule}{\rule{0.5\linewidth}{0.5mm}}
\newcommand{\R}{\mathcal{R}}

\hyphenpenalty=5000
\tolerance=1000

\title{\Large \boldmath \bf Observation of a Neutral Structure near the $D\bar{D}^{*}$ Mass Threshold in $e^{+}e^{-}\to (D \bar{D}^*)^0\pi^0$ at $\sqrt{s}$ = 4.226 and 4.257 GeV}

\author{
  \begin{small}
    \begin{center}
      M.~Ablikim$^{1}$, M.~N.~Achasov$^{9,f}$, X.~C.~Ai$^{1}$,
      O.~Albayrak$^{5}$, M.~Albrecht$^{4}$, D.~J.~Ambrose$^{44}$,
      A.~Amoroso$^{49A,49C}$, F.~F.~An$^{1}$, Q.~An$^{46,a}$,
      J.~Z.~Bai$^{1}$
% ...
      \\
      \vspace{0.2cm}
      (BESIII Collaboration)\\
      \vspace{0.2cm} {\it
        $^{1}$ Institute of High Energy Physics, Beijing 100049, People's Republic of China\\
% ...
$^{4}$ Bochum Ruhr-University, D-44780 Bochum, Germany\\
        $^{5}$ Carnegie Mellon University, Pittsburgh, Pennsylvania 15213, USA\\
% ...
$^{9}$ G.I. Budker Institute of Nuclear Physics SB RAS (BINP), Novosibirsk 630090, Russia\\
% ...
$^{44}$ University of Rochester, Rochester, New York 14627, USA\\
% ...
$^{46}$ University of Science and Technology of China, Hefei 230026, People's Republic of China\\
% ...
$^{49}$ (A)University of Turin, I-10125, Turin, Italy; (B)University of Eastern Piedmont, I-15121, Alessandria, Italy; (C)INFN, I-10125, Turin, Italy\\
% ...
$^{a}$ Also at State Key Laboratory of Particle Detection and Electronics, Beijing 100049, Hefei 230026, People's Republic of China\\
% ...
$^{f}$ Also at the Novosibirsk State University, Novosibirsk, 630090, Russia\\
% ...
      }
    \end{center}
    \vspace{0.4cm}
  \end{small}
}

\affiliation{}

\vspace{0.4cm}


\date{\today}

\begin{abstract}
  \vspace{2cm}
  A neutral structure in the $D\bar{D}^{*}$ system around the $D\bar{D}^{*}$ mass threshold is observed
  with a statistical significance greater than 10$\sigma$ in the
  processes $e^{+}e^{-}\rightarrow D^{+}D^{*-}\pi^{0}+c.c.$ and
  $e^{+}e^{-}\rightarrow D^{0}\bar{D}^{*0}\pi^{0}+c.c.$ at $\sqrt{s}$
  = 4.226 and 4.257 GeV in the BESIII experiment. The structure is
  denoted as $\zcn$. Assuming the presence of a resonance, its pole
  mass and width are determined to be
  ($3885.7^{+4.3}_{-5.7}$(stat)$\pm 8.4$(syst))~MeV/$c^{2}$ and
  ($35^{+11}_{-12}$(stat)$ \pm 15$(syst))~MeV, respectively. The Born
  cross sections are measured to be $\sigma(e^{+}e^{-}\to \zcn\pi^{0}, \zcn \to
  D\bar{D}^{*})=(77 \pm 13$(stat)$\pm 17$(syst)) pb at 4.226 GeV and
  ($47 \pm 9$(stat)$ \pm 10$(syst)) pb at 4.257 GeV. The ratio of decay
  rates $\frac{\br{\zcn \to D^{+}D^{*-}+c.c.}}{\br{\zcn \to
      D^{0}\bar{D}^{*0}+c.c.}}$ is determined to be $0.96 \pm
  0.18$(stat)$\pm 0.12$(syst), consistent with no isospin violation in the process $\zcn\to D\bar{D}^*$.
\end{abstract}


\maketitle


The existence of exotic states beyond those of conventional mesons and
baryons was debated for decades, mostly because no convincing
experimental evidence for them had been
found~\cite{Agashe:2014kda}. In recent years, the discovery of charged
$Z_c$ charmonium-like states~\cite{snowmass2013,theor}, which decay to
a charmonium state plus a pion or a pair of charmed mesons and,
therefore, must consist of at least a four constituent quark
configuration $c\bar{c}q\bar{q}'$, has stirred excitement about
these possible exotic states.  In $e^+e^-\to \pi^{\mp} Z_c^{\pm}$
processes, four $Z_c^{\pm}$ states have been discovered in the decays
of $Z_c(3885)^{\pm}\to
(D\bar{D}^{*})^{\pm}$~\cite{Ablikim:2013xxp,binwang},
$Z_c(3900)^{\pm}\to
\pi^{\pm}J/\psi$~\cite{BESprl110,BELLEprl110,Xiao:2013iha},
$Z_c(4020)^{\pm}\to \pi^{\pm} h_c$~\cite{Ablikim:2013gyp}, and
$Z_c(4025)^{\pm}\to
(D^{*}\bar{D}^{*})^{\pm}$~\cite{Ablikim:2014lxr}. There have been many
theoretical predictions and interpretations~\cite{theor} to explain
their nature as exotic mesons. However, none of these models have
either been ruled out or established experimentally.

After the discoveries of the charged $Z_c^{\pm}$ states, BESIII
reported studies of their neutral partners in the isospin symmetric
channel of $e^+e^-\to \pi^0 Z_c^0$. A $Z_{c}(3900)^{0}$ is found in
$e^{+}e^{-}\rightarrow\pi^{0}\pi^{0}J/\psi$~\cite{Ablikim:3900n}, a
$Z_{c}(4020)^{0}$ in
$e^{+}e^{-}\rightarrow\pi^{0}\pi^{0}h_{c}$~\cite{Ablikim:4020n}, and a
$Z_{c}(4025)^{0}$ in
$e^{+}e^{-}\rightarrow\pi^{0}(D^{*}\bar{D}^{*})^{0}$~\cite{Ablikim:4025n}.
Evidence for $Z_{c}(3900)^{0}$ in $e^+e^-\to \pi^0 Z_c^0$ was
previously reported with CLEO-c data at $\sqrt{s} = 4.17$
GeV~\cite{Xiao:2013iha}.  These measurements indicate that the
$Z_c(3900)$, $Z_c(4020)$ and $Z_c(4025)$ are three different isospin
triplet states, since their relative Born cross sections of the
charged modes to the neutral modes are compatible with isospin
conservation. This motivates a search for the neutral partner of the
$\zcpm$ in $e^{+}e^{-}\to (D\bar{D}^{*})^0\pi^{0}+c.c.$ to identify
its isospin.

In this Letter, the process $e^{+}e^{-}\rightarrow
(D\bar{D}^*)^0\pi^{0}+c.c.$ is studied, where $(D\bar{D}^*)^0$ refers
to $D^{+}D^{*-}$ or $D^{0}\bar{D}^{*0}$. A neutral charmonium-like
structure, the $\zcn$, is observed around the $(D\bar{D}^*)^0$ mass
threshold in the $(D\bar{D}^*)^0$ mass spectrum.  This analysis is
based on data samples collected by the BESIII detector with
integrated luminosities of 1092 pb$^{-1}$ at $\sqrt{s}$ = 4.226 GeV
and 826 pb$^{-1}$ at $\sqrt{s}=$4.257 GeV~\cite{Abli:lumi,Ecms}.  Note that
charge conjugation is always implied, unless explicitly stated.

BESIII~\cite{Ablikim2010345} is a general-purpose detector at the double-ring
$e^+e^-$ collider BEPCII, which is used for the study of physics in
the $\tau$-charm energy region~\cite{taucharm}.
Monte Carlo~(MC) simulations based on {\sc Geant4}~\cite{GEANT4} are implemented in the BESIII experiment. For each energy point, we generate a
signal MC sample based on the covariant tensor amplitude
formalism~\cite{CTAF} to simulate the $S$-wave process $e^{+}e^{-}\to
Z_{c}^{0}\pi^{0}\to(D\bar{D}^{*})^0\pi^{0}$, assuming that the
$Z_{c}^{0}$ has $J^{P}=1^+$. Effects of initial state radiation
are taken into account with the MC event generator {\sc
  kkmc}~\cite{kkmc1,kkmc2}, where the line shape of the Born cross
section of $e^{+}e^{-}\to Z_{c}^{0}\pi^{0}\to(D\bar{D}^{*})^0\pi^{0}$
is assumed to follow that of the charged channel $e^{+}e^{-} \to
Z_{c}^{\pm}\pi^{\mp}\to(D\bar{D}^{*})^{\pm}\pi^{\mp}$~\cite{Ablikim:2013xxp}. In
addition, a large statistics MC sample of the three body process
$e^{+}e^{-} \to (D\bar{D}^{*})^0\pi^{0}$ is generated according to
phase space~(PHSP). To study possible backgrounds, MC simulations of
$Y(4260)$ generic decays, initial state radiation production of the vector charmonium
states, charmed meson production, and the continuum process
$e^{+}e^{-}\to q\bar{q}$ ($q=u,~d,~s$) equivalent to 10~times
the luminosity of
the data at $\sqrt{s}$ = 4.226 and 4.257 GeV are generated. Particle
decays are simulated with {\sc evtgen}~\cite{evtgen,besevtgen} for the
known decay modes with branching fractions set to the world
average~\cite{Agashe:2014kda} and with the {\sc lundcharm}
model~\cite{lund} for the remaining unknown decays.

In this work, we study $e^{+}e^{-} \to D^{+}D^{*-}\pi^{0}$,
$D^{*-}\to\bar{D}^{0}\pi^{-}$ based on the detection of the
$D^+\bar{D}^0$ pair and $e^{+}e^{-} \to D^{0}\bar{D}^{*0}\pi^{0}$,
$\bar{D}^{*0}\to\bar{D}^{0}\pi^{0}$ based on the detection of the
$D^0\bar{D}^0$ pair. The $D\bar{D}$ meson pairs are reconstructed
through five hadronic decay modes $K^{-}\pi^{+}\pi^{+},
K^{-}\pi^{+}\pi^{+}\pi^{0}, K_{S}\pi^{+},
K_{S}\pi^{+}\pi^{0},K_{S}\pi^{+}\pi^{+}\pi^{-}$ for the $D^+$ and
three modes $K^{+}\pi^{-}$, $K^{+}\pi^{-}\pi^{0}$,
$K^{+}\pi^{+}\pi^{+}\pi^{-}$ for the $\bar{D}^{0}$. The primary
$\pi^{0}$, which is produced along with the $D\bar{D}^*$ in the
$e^+e^-$ reaction, is reconstructed from a pair of photons, while the
soft $\pi$ from the $D^{*}$ decay is not required to improve the
detection efficiency.
The $D^+D^-$ mode is not included because of its low rate compared to
$D^0\bar{D}^0$ and $D^+\bar{D}^0$.

In this analysis, the selection criteria in Ref.~\cite{binwang} are
used to identify the $\pi^\pm / K^\pm$, photon, $\pi^{0}$ and $K_{S}$
candidates. The charged-particle tracks in each $D$ candidate are
constrained to a common vertex, except for those from $K_{S}$ decays,
and the $\chi^{2}$ of the vertex fit is required to be less than
100. Each $D$ candidate is required to have its reconstructed
invariant mass in the range (1.840, 1.880) GeV$/c^{2}$. Furthermore, a
mass-constrained kinematic fit (KF) to the nominal $D$ mass is
performed, and the KF chisquare $\chi^2_{D}$ is required to be less
than 100. In case there is more than one $D\bar{D}$ combination in an
event, only the candidate with the minimum sum of
$\chi^2_{D}+\chi^2_{\bar{D}}$ is kept. The $D\bar{D}$ four-momenta
from the mass-constrained KF are used for the further analysis.

The primary $\pi^{0}$ candidates are reconstructed with pairs of
photons which are not used in forming the $D\bar{D}$ mesons, and their
invariant masses $M(\gamma\gamma)$ must be in the range (0.120, 0.150)
GeV$/c^{2}$. To reduce backgrounds and to improve the resolution, a KF
with 2 degrees of freedom (2C) is performed, constraining
$M(\gamma\gamma)$ to the nominal $\pi^0$ mass $m(\pi^0)$ and the
recoil mass of $\pi^0D\bar{D}$, RM$(\pi^{0}D\bar{D})$, to the nominal
$\pi$ mass. The 2C KF chisquare $\chi^{2}_{\rm 2C}(\pi)$ must be less
than 200.  For each $D\bar{D}$ mode, if there is more than one primary
$\pi^0$ candidate, the one with the minimum $\chi^{2}_{\rm 2C}(\pi)$
is retained for further analysis.  For $e^{+}e^{-}\to
D^{0}\bar{D}^{*0}\pi^{0}$ with $\bar{D}^{*0}\to\bar{D}^{0}\pi^{0}$,
the process $e^{+}e^{-}\to D^{0}\bar{D}^{*0}\pi^{0}$ with
$\bar{D}^{*0}\to\bar{D}^{0}\gamma$ is a major background. To reject
this background, we require $\chi^{2}_{\rm 2C}(\pi^{0}) < 60$. We also
perform a similar 2C KF but constrain RM$(\pi^{0}D^{0}\bar{D}^{0})$ to
be zero, which corresponds to the mass of the photon in
$\bar{D}^{*0}\to\bar{D}^{0}\gamma$, and the corresponding fit
chisquare is required to satisfy $\chi^{2}_{\rm 2C}(\gamma)>20$ to
further suppress this background. The fitted four-momentum of the
primary $\pi^0$ is used in the next stage of the analysis.

\begin{figure}[tp!]
\begin{center}\small
  {\includegraphics[width=1.0\linewidth]{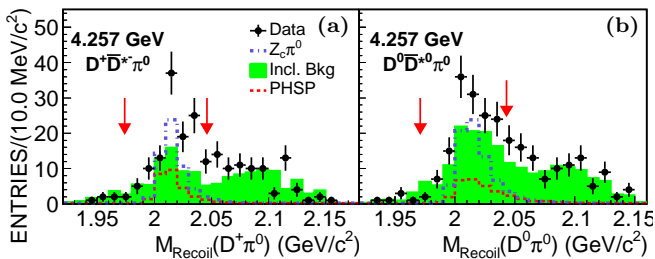}
  \put(-127,88) {\bf \footnotesize (a)}\put(-18,88) {\bf \footnotesize (b)}}
\caption{ Distributions of RM$(D\pi^{0})$ at $\sqrt{s}$ = 4.257
  GeV. The signal and PHSP processes are overlaid with an arbitrary
  scale. The solid arrows indicate the selection criteria for the $(D\bar{D}^*)^{0}\pi^{0}$ candidates. Data at
  $\sqrt{s}$ = 4.226 GeV show similar distributions and are omitted.}
\label{rcMDpi}
\end{center}
\end{figure}

In the surviving events, the occurrence of multiple
$(D\bar{D}^*)^0\pi^0$ combinations per event is
negligible. To help separate the signal events, we require
$M(D^+\pi^{0})>2.1$ GeV/$c^{2}$ and $M(D^0\pi^{0})>2.1$ GeV/$c^{2}$~\cite{supplem}. Because of the limited phase space,
the invariant mass of $D^+\pi^{0}$($D^0\pi^{0}$) and that of
$\bar{D}^0\pi^{0}$ are highly correlated, and the background with the
selected $\pi^0$ and $\bar{D}^0$ from the $\bar{D}^{*0}$ decay is
suppressed by the above selection criteria, too. The RM$(D\pi^{0})$
distributions are illustrated in Fig.~\ref{rcMDpi}, where clear peaks
are seen over simulated backgrounds around the $m(D^{*})$
position. These peaks correspond to the final states of
$(D\bar{D}^{*})^0\pi^{0}$. We further require events to be within the
mass window $|$RM$(D\pi^{0}) - m(D^{*})| <$ 36 MeV/$c^{2}$ for the final
analysis.

\begin{figure}[tp!]
\begin{center}
{\includegraphics[width=1.0\linewidth]{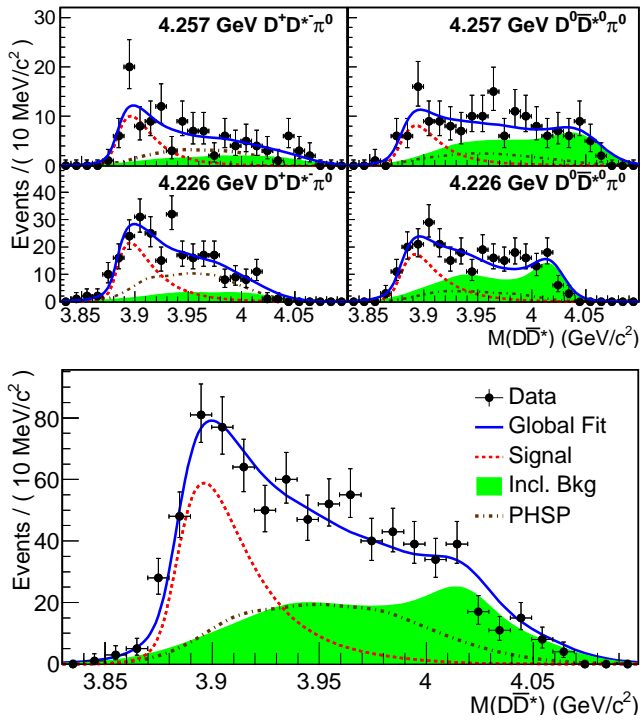}}
\caption{(Upper) Projections of the simultaneous fit to the $M(D\bar{D}^*)$
  spectra for $e^{+}e^{-} \to D^{+}D^{*-}\pi^{0}$ and
  $D^{0}\bar{D}^{*0}\pi^{0}$ at $\sqrt{s}$ = 4.226 and 4.257
  GeV. (Lower) Sum of the simultaneous fit to the $M(D\bar{D}^*)$
  spectra for different decay modes at the different energy points
  above. } \label{simfit}
\end{center}
\end{figure}

The $M(D\bar{D}^*)$ distribution of the surviving events is plotted in
Fig.~\ref{simfit}. An enhancement near the $D\bar{D}^{*}$ mass
threshold around 3.9 GeV/$c^{2}$ is visible, which is seen in both
$D^{+}D^{*-}\pi^{0}$ and $D^{0}\bar{D}^{*0}\pi^{0}$ at $\sqrt{s}$ =
4.226 and 4.257 GeV. As verified in MC simulations, these structures
cannot be attributed to the $e^{+}e^{-}\to(D\bar{D}^{*})^0\pi^{0}$
three body PHSP or inclusive MC background. Possible backgrounds from
$e^{+}e^{-}\to D^{(*)}\bar{D}^{**}\to D\bar{D}^{*}\pi$ have been
studied. Most of them, such as $D^{*}\bar{D}^{*}(2400)$,
$D\bar{D}^{*}(2460)$ and $D^{*}\bar{D}^{*}(2420)$ cannot contribute
to the selected events since their mass thresholds
are higher than 4.26 GeV/$c^{2}$. The only possible peaking background
$e^{+}e^{-}\to D^{(*)}\bar{D}_1(2420)$ has been studied in
Ref.~\cite{binwang}, and its contribution is found to be negligible.

Assuming that there is a resonant structure close to the
$D\bar{D}^{*}$ mass threshold (labeled as $\zcn$), we model its line
shape using a relativistic $S$-wave Breit-Wigner function with a
mass-dependent width multiplied with a phase space factor $q$
$$\bigg| \frac{\sqrt{M\Gamma_{I}(M)/c^{2}}}{M^{2}-m^{2}+iM(\Gamma_{1}(M)+\Gamma_{2}(M))/c^{2}} \bigg|^2 q ~~~ (I=1,2),$$
where $\Gamma_{I}(M) =
\Gamma_{I}\cdot(m/M)\cdot(p^{*}_{I}/p^{0}_{I})$.  $I$ denotes the
different decay modes, where $I=1$ represents the $D^{+}D^{*-}$ decay mode
and $I=2$ represents the $D^{0}\bar{D}^{*0}$ decay mode. $M$ is the reconstructed mass, $m$ is
the nominal resonance mass and $\Gamma_I$ is the partial width of the
decay channel $I$. Under the assumption of isospin symmetry, we take
$\Gamma_I$ to be half of the full width $\Gamma$, assuming that the
decay rates to other possible coupled channels are negligible. $p^{*}_{I}$($q$) is the
momentum of the $D$($\pi^0$) in the rest frame of the $D\bar{D}^{*}$
system (the initial $e^+e^-$ system), and $p^{0}_{I}$ is the momentum
of the $D$ in the resonance rest frame at $M=m$.

An unbinned maximum likelihood fit is performed on the $M(D\bar{D}^*)$
spectra for $e^{+}e^{-} \to (D\bar{D}^{*})^0\pi^{0}$ simultaneously at
$\sqrt{s}$ = 4.226 and 4.257 GeV.  Three components are included in
the fits: the $\zcn$ signal, the PHSP processes and MC simulated
backgrounds.  The signal shape is described as a
mass-dependent-efficiency weighted Breit-Wigner function, described
above, convoluted with the experimental resolution function.  The
resolution function and the efficiency shape are obtained from MC
simulations.  The shape of the PHSP processes is derived from MC
simulations, and their amplitudes are allowed to vary in the fits.
The inclusive MC background distributions are modeled based on the
kernel estimation~\cite{Cranmer}, and their sizes are fixed according
to the expected numbers estimated in the inclusive MC samples. The
simulated backgrounds are validated by comparing their $M(D\pi^0)$ and
RM$(D\pi^{0})$ distributions with those for data in sideband regions
$(1.920, 1.974)\cup(2.090,2.180)$ GeV$/c^{2}$ for the $D^+\bar{D}^0$ mode
and $(1.920, 1.971)\cup(2.090,2.160)$ GeV$/c^{2}$ for the $D^0\bar{D}^0$
mode.

We define the ratio
$\R=\Br_{D^{+}D^{*-}}/\Br_{D^{0}\bar{D}^{*0}}$, where
$\Br_{D^{+}D^{*-}}$($\Br_{D^{0}\bar{D}^{*0}}$) is the branching ratio
of $\zcn \rightarrow D^{+}D^{*-}$($D^{0}\bar{D}^{*0}$). In the fit, $\R$ is assumed to be same for the data at $\sqrt{s}$ = 4.226 and
4.257 GeV. The number of observed signal events, $N_{\rm obs}$, is given by $ N_{\rm obs}
=
\mathcal{L}\sigma_{D\bar{D}^*}(1+\delta^{\rm rad})(1+\delta^{\rm vac})\varepsilon
\Br_{\rm int}$, where $\sigma_{D\bar{D}^*}$ is the Born cross section
$\sigma(e^{+}e^{-} \rightarrow \zcn\pi^{0}, \zcn\to D\bar{D}^{*})$,
$\mathcal{L}$ is the integrated luminosity, (1+$\delta^{\rm rad}$) is the
initial radiative correction factor, (1+$\delta^{\rm vac}$) is the vacuum
polarization factor~\cite{vac}, $\varepsilon$ is the detection
efficiency and $\Br_{\rm int}$ is the product of the decay rates of
the intermediate states.

Figure~\ref{simfit} shows the fit results. To assess the goodness of fit, we bin the data set in 19 bins such that
each bin contains at least 10 events, and compute the $\chi^2$ between
the binned data and the projection of the fit. We find
$\chi^2/\mathrm{d.o.f.} = 18.5/19$ for the simultaneous fit in the lower plot. The statistical significance of the
$\zcn$ signal is estimated to be more than $12\sigma$, based on the difference
of the maximized likelihoods between the fit with and without
including the signal component. The mass and width of the $\zcn$ are
measured to be $m(\zcn) = (3894.7\pm 3.0)$ MeV/$c^{2}$ and
$\Gamma(\zcn) = (36 \pm 17)$ MeV. The corresponding pole mass and
width are calculated to be $m_{\rm pole}(\zcn) = 3885.7^{+4.3}_{-5.7}$
MeV/$c^{2}$ and $\Gamma_{\rm pole}(\zcn) = 35 ^{+11}_{-12}$
MeV~\cite{CWBW}. From the fit, we determine $\sigma_{D\bar{D}^{*}}$ to
be $(77 \pm 13)$ pb and $(47 \pm 9)$ pb at $\sqrt{s}$ = 4.226 and
4.257 GeV, respectively.  We also obtain $\R=0.96 \pm 0.18$.

The systematic uncertainties on the measurements of the $\zcn$ resonance
parameters, the cross section $\sigma_{D\bar{D}^*}$ and the ratio $\R$
are studied, and the major contributions are summarized in
Table~\ref{sumerr}. The systematic uncertainties on the $\zcn$ resonance
parameters mainly come from the signal shape, background, mass shift
and detector resolution. The dominant systematic uncertainties on
$\sigma_{D\bar{D}^*}$ and $\R$ are from the background, resolution and
detection efficiency.

\begin {table}[tp!]
\centering
\small
\caption{Summary of systematic uncertainties for the resonance parameters, the Born cross sections and the ratio of decay
rates. Values outside the parenthesis represents uncertainties for
  $\sigma_{D\bar{D}^*}$ at $\sqrt{s}$ = 4.226 GeV, while those inside
  are for $\sigma_{D\bar{D}^*}$ at $\sqrt{s}$ = 4.257 GeV. The total
  systematic uncertainties are obtained by combining all the
  independent sources in quadrature. }\label{sumerr}
\footnotesize
\begin {tabular}{l|c|c|c|c}
\hline \hline
Source                              & $m_{\rm pole}$(MeV/$c^{2}$)     &  $\Gamma_{\rm pole}$(MeV)       & $\sigma_{D\bar{D}^*}(\%)$ & $\R(\%)$    \\
\hline
Beam energy                          &  1.0               &   3.0                 & 4  (5)   & 1     \\
Signal shape                        &  3.5               &   8.2                 & 5  (4)   & 2     \\
Background                          &  6.8               &   6.6                 &15  (15)  & 4      \\
Fit range                           &  0.3               &   0.3                 & 3   (1)   & 1     \\
Mass shift                          &  3.0               &                        &           &\\
Resolution                          &                    &   9.5                  &11   (4)  & 1     \\
Efficiency                          &                    &                        &11     (11)   & 11     \\
Input-output check                  &  1.6               &   2.5                  &              &        \\
(1+$\delta^{\rm rad}$)(1+$\delta^{\rm vac}$)      &                    &                        & 5    (5)   &      \\
$\mathcal{B}_{\rm int}$                     &                    &                        & 5     (5)   & 5      \\
$\mathcal{L}$                      &                       &                         & 1 (1)   & \\ \hline
Total                               &  8.4               &  15                 & 23     (21)   &13 \\
\hline
\end {tabular}
\end {table}

The uncertainty from the beam energy is estimated by varying the beam
energy by $\pm 1$ MeV in the 2C KF, and the maximum differences of the
mass, width, $\sigma_{D\bar{D}^*}$ at $\sqrt{s}$ =4.226 (4.257) GeV
and $\R$ are found to be 1.0 MeV/$c^{2}$, 3.0 MeV, 5\%(4\%) and 1\%,
respectively.  To assess the uncertainty of the signal shape, an
$S$-wave relativistic Breit-Wigner function with constant
width~\cite{CWBW} is taken as an alternative signal model in the simultaneous
fit. The changes of the fitted mass and width are determined to be 3.5
MeV/$c^{2}$ and 8.2 MeV, while the change on $\sigma_{D\bar{D}^*}$ is
5\%(4\%) at $\sqrt{s}$ =4.226 (4.257) GeV and on $\R$ 2\%.  The
systematic uncertainty due to the background description is estimated by
leaving free the absolute numbers of the inclusive backgrounds in the fit,
or adjusting their shapes by varying the scalings of different
background components in the inclusive MC samples. Those fit results
differ from the nominal results by 6.8 MeV/$c^{2}$ in mass, 6.6 MeV in
width, 15\% in $\sigma_{D\bar{D}^*}$ both at $\sqrt{s}$ =4.226 and
4.257 GeV, and 4\% in $\R$. Maximum fluctuations due to changing the
fit range are assigned as systematic uncertainties. The MC simulation
of the mass shift and resolution may not fully reflect the effects in
data, and it is studied by fitting the $\bar{D}^*$ peak in the
RM$(D\pi^{0})$ spectra to obtain the mass shift and the resolution
difference between data and MC simulations. The obtained mass shift is quoted as
part of the systematic uncertainties of the mass. The variations of the
fit results after considering the resolution difference is
assigned as systematic uncertainty.

Efficiency-related systematic uncertainties are universal in each $D$
decay mode and include six sources: tracking efficiency, particle
identification, photon detection efficiency, $\pi^{0}$ reconstruction
efficiency, $K_{S}$ reconstruction efficiency and KF efficiency. The
uncertainties of tracking efficiency and particle identification for
$\pi^\pm$ and $K^\pm$ are evaluated to be 1\% per
track~\cite{k-eff,pi-eff}.  The uncertainty in the
photon-reconstruction efficiency is estimated to be about 1\% per
photon~\cite{pho-eff}. The efficiency difference of reconstructing the
$K_S$ in MC simulations and in data is 4.0\%~\cite{Ks}. The
uncertainty in $\pi^{0}$ reconstruction is 1\%~\cite{pho-eff}. The
systematic bias of the KF is estimated by using the
track-parameter-correction method~\cite{kmfit}. The correction factors
for helix track parameters are determined from the control sample
$e^{+}e^{-} \to K^{*}(892)^0K^+\pi^- \rightarrow
K^{+}K^{-}\pi^{+}\pi^{-}$. The total efficiency-related systematic
uncertainty is taken as the square root of the quadratic sum of the
individual uncertainties. The potential bias from the event selection
and the analysis procedure is studied with input-output
checks, which compare the output results with the input values of the
resonance mass and width based on MC simulations.  We assign the
systematic uncertainty of 1.6 MeV/$c^{2}$ in mass and 2.5 MeV in width
accordingly. The systematic uncertainty of the radiative correction
factor $1+\delta^{\rm rad}$, which includes the effect on the
detection efficiency, is estimated to be 5\% by changing the input
$(D\bar{D}^{*})^0\pi^{0}$ line shape within
errors~\cite{Ablikim:2013xxp}. The systematic uncertainty of the
vacuum polarization factor $1+\delta^{\rm vac}$ is 0.5\% taken from
the QED calculation~\cite{vac}.  The weighted systematic uncertainty of
$\mathcal{B}_{\rm int}$ is from the world average
value~\cite{Agashe:2014kda}.  The uncertainty of integrated luminosity
is taken as 1\% by measuring Bhabha events~\cite{Abli:lumi}. The
uncertainty of the mass window requirement is negligible. The overall
systematic uncertainties are determined by combining all the sources
in quadrature, assuming they are independent.

In summary, we study $e^{+}e^{-}\rightarrow
D^{+}D^{*-}\pi^{0}+c.c.$ and $e^{+}e^{-}\rightarrow
D^{0}\bar{D}^{*0}\pi^{0}+c.c.$ using data taken at $\sqrt{s}=4.226$
and 4.257 GeV. A neutral structure around the $D\bar{D}^{*}$ mass
threshold is observed with a statistical significance greater than
$10\sigma$. Assuming that it is a resonance, we model it with a relativistic
Breit-Wigner function. Its pole mass and width are measured to be
($3885.7 ^{+4.3}_{-5.7}$(stat)$\pm 8.4$(syst)) MeV/$c^{2}$ and ($35
^{+11}_{-12}$(stat)$\pm 15$(syst)) MeV, respectively, which are close
to the mass and width of the reported charged
$\zcc$~\cite{Ablikim:2013xxp,binwang}. The Born cross sections
$\sigma(e^{+}e^{-}\to Z_{c}^{0}\pi^{0}\to(D\bar{D}^{*})^0\pi^0 +
c.c.)$ are determined to be $(77 \pm 13 \pm 17)$ pb and $(47 \pm 9 \pm
10)$ pb at $\sqrt{s}=4.226$ and 4.257 GeV, respectively, which are
consistent with half of $\sigma(e^{+}e^{-} \rightarrow
Z_{c}^{+}\pi^{-} \rightarrow
(D\bar{D}^{*})^{+}\pi^{-}+c.c.)$~\cite{binwang}. A comparison between the resonance parameters of the $\zcc$ and the $\zcn$ is summarized in the Supplemental Material~\cite{supplem}. All these observations favor the
assumption that the $\zcn$ is the
neutral isospin partner of the $\zcpm$, and the $\zcpm$/$\zcn$ form an isospin
triplet. In addition, we determine the ratio
of the decay rate $\R = \frac{\br{\zcn \to
    D^{+}D^{*-}}}{\br{\zcn \rightarrow D^{0}\bar{D}^{*0}}} = 0.96\pm
0.18 \pm 0.12$, which is consistent with unity. Hence, no isospin violation in the process $\zcn\to D\bar{D}^*$ is
observed.

\begin{acknowledgements}
The BESIII collaboration thanks the staff of BEPCII and the IHEP computing center for their strong support. This work is supported in part by National Key Basic Research Program of China under Contract No.~2015CB856700; National Natural Science Foundation of China (NSFC) under Contracts Nos.~11125525, 11235011, 11322544, 11335008, 11425524; the Chinese Academy of Sciences (CAS) Large-Scale Scientific Facility Program; the CAS Center for Excellence in Particle Physics (CCEPP); the Collaborative Innovation Center for Particles and Interactions (CICPI); Joint Large-Scale Scientific Facility Funds of the NSFC and CAS under Contracts Nos.~11179007, U1232201, U1332201; CAS under Contracts Nos.~KJCX2-YW-N29, KJCX2-YW-N45; 100 Talents Program of CAS; National 1000 Talents Program of China; INPAC and Shanghai Key Laboratory for Particle Physics and Cosmology; German Research Foundation DFG under Contract No.~Collaborative Research Center CRC-1044; Istituto Nazionale di Fisica Nucleare, Italy; Ministry of Development of Turkey under Contract No.~DPT2006K-120470; Russian Foundation for Basic Research under Contract No.~14-07-91152; The Swedish Resarch Council; U.S.~Department of Energy under Contracts Nos.~DE-FG02-04ER41291, DE-FG02-05ER41374, DE-SC0012069, DESC0010118; U.S.~National Science Foundation; University of Groningen (RuG) and the Helmholtzzentrum fuer Schwerionenforschung GmbH (GSI), Darmstadt; WCU Program of National Research Foundation of Korea under Contract No.~R32-2008-000-10155-0.
\end{acknowledgements}

\end{document}